\documentstyle[prd,aps,preprint,epsfig]{revtex}
\tighten

\begin{document}
\draft
 
\pagestyle{empty}

\preprint{
\noindent
%\begin{minipage}[t]{3in}
%\begin{flushleft}
%\today \\
%\end{flushleft}
%\end{minipage}
\hfill
\begin{minipage}[t]{3in}
\begin{flushright}
LBNL--42025 \\
%UCB--PTH--98/xx \\
%hep-ph/98xxxxx \\
July 1998
\end{flushright}
\end{minipage}
}

\title{The final-state interaction in the two-body nonleptonic decay
of a heavy particle}

\author{
Mahiko Suzuki
%\thanks{Work supported in part by the Director, Office of Energy
%Research, Office of High Energy and Nuclear Physics, Division of High
%Energy Physics of the U.S. Department of Energy under Contract
%DE--AC03--76SF00098 and in part by the National Science Foundation under
%grant PHY--95--14797.}
}
\address{
Department of Physics and Lawrence Berkeley National Laboratory\\
University of California, Berkeley, California 94720
}

%\thanks{Work supported by the Department of Energy under Contract
%DE--AC03--76SF00515.}

%\date{\today}
\maketitle

\begin{abstract}
We attempt to understand the final-state interaction in the two-body 
nonleptonic decay of a heavy particle for which many multibody 
($N\geq 3$) decay channels are also open. No matter how many multibody  
channels couple to the two-body channels, the analyticity of the S-matrix 
relates the phase and the magnitude of the two-body decay amplitude 
through a dispersion relation. In general, however, the phase cannot be 
determined by strong interactions alone. The dispersion relation requires 
on a general ground that the final-state interaction phases be small for 
the two-body decay amplitudes when the initial particle is very heavy.
We then analyze the final-state interaction phases in terms of the 
$s$-channel eigenstates of the S-matrix and obtain semiquantitative 
results applicable to the $B$ decay with a random S-matrix hypothesis. 
We use the high-energy scattering data and the dual resonance model
as a guide to the relevant aspects of strong interaction dynamics 
at long and intermediate distances. 

\end{abstract}
\pacs{PACS numbers: 11.55.Fv, 11.80.Gw, 12.40.Nn, 13.20.Fc, 13.20.He}
%\newpage
\pagestyle{plain}
\narrowtext

\setcounter{footnote}{0}

\section{Introduction}

The final-state interaction in the nonleptonic weak decay is difficult 
to estimate when a large number of multibody channels are open. 
While the short-distance final-state interaction is small\cite{SD} 
and its computation is noncontroversial, we have little understanding, 
theoretically or experimentally, of the long-distance final-state 
interactions.

The long-distance final-state interaction phases were computed from 
the high-energy Regge exchange amplitudes in the elastic rescattering 
approximation\cite{Zhang,KL}. The experimental data appeared in
favor of large final-state interaction phases at least for the $D$
decay\cite{Bishai}. It was asserted that the measured phases 
of the two-body $D$ decay amplitudes can be reproduced in the
elastic approximation to the final hadron interactions\cite{Weyers}.
However, it is fairly obvious from an analysis of the partial-wave 
unitarity with the diffractive scattering\cite{Donoghue} that the 
elastic approximation cannot be justified at high mass scales,
for instance, in the B decay where the two-body final states 
can couple to a very large number of multibody final states. Even 
when we are interested in the final-state interaction phases of the 
two-body channels alone, we cannot determine them without knowing the 
coupling of the two-body channels to the multibody channels. The Regge 
amplitudes alone do not provide all necessary pieces of information. 
Actually, strong interactions and CP-conserving weak interactions are
entangled in the decay phases. In this paper we shall make a modest 
attempt toward understanding of the inelastic final-state interactions. 
Because of the limitation in the numerical computation of the 
long-distance effects, we are able to present our results only in 
a semiquantitative way.

  We present two approaches here. The first one uses the analyticity of the
S-matrix. For the two-body decay, the phase and the magnitude of amplitude
are tightly related to each other by a dispersion relation no matter how 
many multibody channels couple to two-body channels. The same phase-amplitude
relation does not hold for the multibody decay. So far no theorist has 
ever attempted to study the correlation between the phase and the magnitude 
from this aspect. No dynamical assumption is introduced in this approach.  

The origin of difficulty in the final-state interaction at high mass 
scales is in that so many channels are open and communicate with each other.
In the second approach we analyze the decay in terms of the $s$-channel
eigenstates of the S-matrix and treat a large number of open eigenchannels 
statistically by introducing a randomness hypothesis\cite{CS}. 
While we give up much of numerical predictability in this approach, 
we are still able to see general trends in the final-state interaction at 
high mass scales. Both approaches lead us to conclude that the 
long-distance final-state interaction phases should be small for 
two-body heavy hadron decay. Though our conclusion favoring small 
final-state interactions may be in line of some of the existing literature, 
our method and picture are completely orthogonal to them.

In Section II, after a brief review of the analyticity of the decay 
amplitude into general $N$-body channels, we derive for the two-body decay 
a dispersion relation which relates the phase and the amplitude through 
the Omn\`{e}s-Mushkelishvili integral\cite{OM}. Using this dispersion relation 
we separate from the physical decay amplitude the final-state interactions 
below an arbitrarily chosen timelike energy scale. We see in this form
that a final-state interaction phase of any origin cannot persist to 
very high energies. 
     
In Section III, we shall study the final-state interaction phases 
from the viewpoint of the eigenphase shifts of the strong-interaction 
S-matrix. We write the hadron scattering amplitude in terms of 
the eigenphase shifts, and classify the eigenchannels into the resonant 
and nonresonant ones according to the dual resonance model. Then we 
express the final-state interaction in terms of the eigenphase shifts. 
We study the high-energy behaviors of the eigenphase shifts from the 
scattering data. 

In Section IV we introduce the dynamical postulate that the composition 
of eigenchannels is statistically random when very many of them exist 
in degeneracy. We make a quantitative estimate of the decay amplitude 
phases within the limitation of the method.

Finally in Section V, we apply our findings to the actual $D$ and $B$ 
decays. For the $D$ decay, the elastic approximation combined with 
the Regge asymptotic behavior is not allowed in determining the final-state 
interaction phases. The random-phase method is probably a poor approximation.  
The observed large phase difference between different isospin channels 
can be accommodated but not predicted. The random phase approximation 
has the best chance in the $B$ decay where the final hadron multiplicity 
is high. All two-body decay amplitudes are dominantly real up to CP 
violations in the $B$ decay. If the color suppression exists prior to 
final-state interaction corrections, it should be preserved even in the
presence of final-state interactions. Interestingly, the same high-energy 
behavior of the eigenphase shifts that makes the elastic scattering 
amplitudes purely imaginary leads to the almost real two-body decay 
amplitudes.

\section{Basic properties of the inelastic decay amplitudes}

\subsection{Analyticity}

  We consider the weak decay 
\begin{equation}
                   H\rightarrow{\rm hadrons}
\end{equation}
where $H$ is a heavy particle such as the $D$ and $B$ mesons. 
The final state is generally a multiparticle state of $N$ hadrons.  
Going off the $H$ mass shell, we call the $H$ mass squared
as the variable $s$ and examine the analytic property of
the invariant decay amplitude $M(s)$ in the complex $s$-plane.

The analytic property of the S-matrix elements was extensively studied  
decades ago\cite{Eden}. We obtained many rules of computation by examining 
the Feynman diagrams though a rigorous proof without referring to the 
diagrams was given only in the limited cases. To avoid inessential 
complications, we consider the case where $H$ and the final hadrons are 
all spinless. We shall use the {\it in-out} formalism\cite{LSZ} 
to simplify our notations. The invariant amplitude for the decay into the 
$N$-hadron state $f$ is defined by
\begin{equation}
 \langle f^{out}(p_i)|H_w|H(P)\rangle = M^{+}(p_i),\;\;\;(\sum_ip_i=P)
            \label{amp}
\end{equation}
where the one-particle states are normalized as $\langle {\bf p}|{\bf p}'
\rangle = (2\pi)^3\delta({\bf p}-{\bf p}')$. $M^+(p_i)$ is actually the 
function of all possible Lorentz invariants made of $p_i$ and $P$. 
Aside from $s (=P^2)$, there are $(N+1)(N-2)/2$ independent
invariants, which we may choose as
\begin{equation}
          s_{ij} = (p_i + p_j)^2 \;\;\; (i>j(\neq N-1)),
\end{equation}
since there is one linear dependency relation,
\begin{equation}
           s = 2\sum_{i>j}(p_i\cdot p_j) + \sum_i m_i^2.
\end{equation}
We may include short-distance strong interactions in $H_w$ by using the 
QCD-corrected effective Hamiltonian. Though we shall treat $H_w$ as 
a local operator, the analytic property to be discussed below does not 
depend on locality of $H_w$. What is important is that $H_w$ transfers 
no energy-momentum.

To study analyticity, we introduce the auxiliary (unphysical) amplitude,
\begin{equation}
 \langle f^{in}(p_i)|H_w|H(P)\rangle = M^{-}(p_i).
\end{equation}
We can obtain $M^{-}(p_i)$ by imposing the incoming boundary condition
on the $N$ hadrons in Eq.(\ref{amp}).  Diagramatically, it amounts to 
flipping the sign of $i\epsilon$ in all Feynman propagators. If $H_w$ is 
time-reversal invariant, it holds that 
\begin{equation}
         \langle f^{out}(p_i)|H_w|H(P)\rangle 
     = \langle H(P')|H_w|f(p'_i)^{in}\rangle, \label{timereversal}
\end{equation} 
where $p'_i$ and $P'$ are obtained from $p_i$ and $P$ by reversing the
signs of their space components. Eq.(\ref{timereversal}) reads 
\begin{equation}
            M^{+}(p_i) = M^{-}(p'_i)^*,   \label{timereversal2}
\end{equation}
where the asterisk denotes a complex conjugate. The reversal of the
signs of the space components of $p_i$ does not change the Lorentz
invariants, $s_{ij}$ and $s$. Since both $M^{+}(p_i)$ and $M^{-}(p'_i)$ 
are real below all thresholds and therefore coincide with 
each other there, Eq.(\ref{timereversal2}) means 
that they actually represent values of a single analytic function 
on two different Riemannn sheets. In terms of the Lorentz invariants, 
Eq.(\ref{timereversal2}) can be written as
\begin{equation}
            M(s+i\epsilon, s_{ij}+i\epsilon)
                  = M(s-i\epsilon, s_{ij}-i\epsilon)^*. \label{RA}
\end{equation}
When $H_w$ is not T-invariant, it is convenient to work with each  
short-distance-corrected weak Hamiltonian individually after separating 
out a T-violating phase; $H_w=\overline{H}_we^{i\delta_{CP}}$. Then 
Eq.(\ref{RA}) is valid for $\overline{H}_w$. We shall use the words 
"T-violation" and "CP-violation" as equivalent, assuming the CPT invariance.

One problem about the multibody decay amplitude is that there are 
so many variables; three independent variables even for the three-body 
decay, one more than in the Mandelstam representation for two-body 
scattering.  More a serious obstacle is that the real analyticity relation
Eq.(\ref{RA}) holds only when we go across all cuts in $s$ and $s_{ij}$ 
from $+i\epsilon$ to $-i\epsilon$ simultaneously. We are able to write 
a dispersion relation in one of the variables keeping the others fixed, 
for instance, in variable $s$ keeping $s_{ij} ((ij)\neq(N,N-1)$)
fixed to $s_{ij}+i\epsilon$.  Then $M(s-i\epsilon, s_{ij}+i\epsilon)$,
which is not simply related to the physical amplitude, enters the dispersion 
integral. In the case of a three-body final state, for instance, such 
an amplitude is the complex conjugate of the unphysical amplitude 
$M(s+i\epsilon, s_{ij}-i\epsilon)$ in which the particle pairs 
(1,2) and (1,3) interact with the wrong sign phases $-\delta_{12}$ and 
$-\delta_{13}$, respectively, while the particle pair (2,3) interacts 
with the right sign phase $\delta_{23}$. Only for the two-body decay 
amplitudes does Eq.(\ref{RA}) give the simple real analyticity relation:
\begin{equation}
       M(s+i\epsilon) = M(s-i\epsilon)^*,
\end{equation}
so that the discontinuity in the complex $s$-plane is directly related 
to a physical process.  For the decays of $N\geq 3$, we can write only 
multivariable dispersion relations as an extension of the Mandelstam
representation. We see little chance of extracting a useful information 
out of them. 

\subsection{Dispersion relation}

The standard dispersion relation relates the real part of $M(s)$ to 
the imaginary part. To relate the phase to the magnitude, we write the 
dispersion relation for logarithm of $M(s)$. Notice that $\ln M(s)$ 
has the same analytic property as $M(s)$ except for cuts due to zeros 
of $M(s)$. Choosing the contour of the Cauchy integral as usual, we can 
write the dispersion relation in the once-subtracted form:
\begin{equation}
     \ln M(s)-\ln M(0)= \frac{s}{\pi}\int_{s_0}^{\infty}
                 \frac{\delta(s')}{s'(s'-s)}ds', \label{Omnes1}
\end{equation}
where $s_0$ is the lowest threshold of final states. The phase 
$\delta(s)$ should be normalized to zero at $s=s_0$ in this representation 
to keep $\ln M(s)$ finite at $s=s_0$. When Eq.(\ref{Omnes1}) is 
exponentiated, it is the representation of Mushkelishvili which was
first applied by Omn\`{e}s to a study of the electromagnetic form factors. 
Hereafter we shall refer to this exponentiated dispersion relation as the 
Omn\`{e}s-Mushkelishvili representation\cite{OM}. The application was limited 
to the energy region where the two-body scattering is elastic. Contrary to 
some misconception\cite{KL}, however, the same dispersion relation 
can be derived for the two-body decay even in the presence of inelastic 
channels. We emphasize this point since otherwise the representation 
will be of no use in the heavy hadron decay. The key observation here 
is that once the phase along the cut is known, a real analytic function 
is unique. 

Actually there is one uncertainty that cannot be fixed by the analyticity.  
It arises from possible zeros of $M(s)$. If $M(s)$ has zeros at $s_i$ 
($i=1,2,3,\cdots$), they generate logarithmic singularities for 
$\ln M(s)$ and contribute to the dispersion integral.  When such zeros are
included, the amplitude $M(s)$ is expressed as
\begin{equation}
     M(s) = P(s)\exp\biggl(\frac{s}{\pi}\int_{s_0}^{\infty}
      \frac{\delta(s')}{s'(s'-s)}ds'\biggr),  \label{Omnes2}
\end{equation}
where $P(s)=M(0)\Pi_i(1-s/s_i)$.  We need some physical argument to 
determine the polynomial $P(s)$. In the days of the bootstrap theory 
we used to resort to a certain philosophy of determinism: there must not 
be a free parameter which we cannot control on. We do not think that 
we can argue along the same line in the context of QCD. In our case 
we avoid this problem as follows. 

Let us write Eq.(\ref{Omnes2}) back in the form
\begin{equation}
 \ln M(s)  = \ln P(s)+\lim_{\Lambda^2\rightarrow\infty}\frac{1}{\pi}
 \biggl(\int_{s_0}^{\Lambda^2}\frac{\delta(s')}{s'-s}ds'-
            \int_{s_0}^{\Lambda^2}\frac{\delta(s')}{s'}ds' \biggr)       
\end{equation}
and then define $M(s;m^{*2})$ by
\begin{equation}
 \ln M(s;m^{*2})  = \ln P(s) 
     + \lim_{\Lambda^2\rightarrow\infty}\frac{1}{\pi}
     \biggl(\int_{m^{*2}}^{\Lambda^2}\frac{\delta(s')}{s'-s}ds'-
            \int_{s_0}^{\Lambda^2}\frac{\delta(s')}{s'}ds' \biggr).
\end{equation}
$M(s;m^{*2})$ is the unphysical decay amplitude in which the final-state
interactions of the energy range from the threshold to $\sqrt{s} = m^*$ 
has been removed. In terms of this amplitude, the physical decay matrix 
element $M(s)$ is expressed as
\begin{equation}
       M(s) =  M(s;m^{*2})\exp\biggl(\frac{1}{\pi}\int_{s_0}^{m^{*2}}
      \frac{\delta(s')}{s'-s}ds' \biggr). \label{Omnes3}
\end{equation}  
In this form all strong-interaction corrections below $\sqrt{s} = m^*$ 
are explicitly factored out in the exponent. The correction factor 
includes both short- and long-distance corrections except for the 
interaction responsible for formation of hadrons. Concerning the hadron 
formation interaction, we encounter a fundamental issue in the 
final-state interaction theory. Both the final-state interactions and the 
formation of hadrons result from the same QCD force.  Nonetheless, in
order to formulate the final-state interaction theory, we must separate
the hadron formation forces from the long-distance interactions 
between hadrons.\footnote{Recall that the hadrons were considered 
as elementary when the phase theorem was proved\cite{Watson}.}
The representation above provides one way to do so.  
While $M(s;m^{*2})$ cannot be measured in experiment, 
it is the closest to what theorists have been calculating for decay 
matrix elements by various methods without final-state interactions. 
The appropriate choice of a value for $m^*$ in Eq.(\ref{Omnes3}) is
\begin{equation}
                m_H \leq m^* \leq m_W.
\end{equation}

A remark is in order on the two-body final state. The dispersion 
relation in Eq.(\ref{Omnes3}) holds for any two-body final 
state. In the elastic energy region, we normally choose an isospin 
or an SU(3) eigenstate for it.  The reason is that such 
a state is an eigenchannel of the strong interaction S-matrix, 
and therefore that the phase is identified with the strong interaction 
phase by Watson's theorem\cite{Watson}. At high energies where 
two-body states couple to multibody final states, however, 
two-body final states are no longer eigenstates of the S-matrix,  
no matter which isospin states we may choose. Then it happens that
the net phase $\delta(s)$ of the two-body decay amplitude depends
not only on strong interactions but also on weak interactions, even 
when one computes amplitudes due to a single effective weak Hamiltonian.
Though this was already pointed out\cite{Donoghue} in the past,
it is worth emphasizing since it is the origin of all complications 
when rescattering is inelastic. We shall see the point more clearly in 
Section III. From this viewpoint, two-body isospin eigenstates in 
the inelastic region are just as bad as the $I_3$ eigenstates of 
indefinite isospin in the elastic region. When inelastic channels
are open, writing for instance the decay amplitudes for 
$B^0\rightarrow D^-\pi^+, \rightarrow\overline{D}^0\pi^0$ 
in the $I=1/2$ and $3/2$ amplitudes contributes to very 
little to solving the problem.

\subsection{High-energy behavior of the phase}

Imagine that $\delta(s)$ approaches the asymptotic value 
$\delta(\infty)$ at some value $m$ well below $\sqrt{s} = 
m_H$:\footnote{Since there is no large characteristic energy scale 
of strong interactions, it is reasonable to assume that $M(s)$ 
approaches its asymptopia as early as hadron scattering amplitudes do. 
If $\delta(s)$ should keep oscillating, $\ln M(s)$ would behave 
like $e^{|s|}$ along some direction in the complex $s$-plane, 
which would prevent us from writing the dispersion relation to 
start with.}
\begin{equation}
   \delta(s) \simeq \delta(\infty) \;\;\;
        (m \leq \sqrt{s} \leq m_W). \label{asym}
\end{equation} 
The exponent of Eq.(\ref{Omnes3}) can be estimated for such $\delta(s)$ as
\begin{eqnarray}
 \frac{1}{\pi}\int_{s_0}^{m^{*2}}\frac{\delta(s')}{s'-m_H^2-i\epsilon}ds'
        &=& \frac{1}{\pi}\biggl(\int_{m^2}^{m^{*2}}+\int_{s_0}^{m^2}\biggr)
              \frac{\delta(s')}{s'-m_H^2}ds', \\ \nonumber
        &=& \frac{2\delta(\infty)}{\pi}\ln\biggl(\frac{m*}{m_H}\biggr)
          + i\pi +{\rm O}\biggl(\frac{m^2}{m_H^2}\biggr).  \label{integral}
\end{eqnarray}
The contribution of O$(m^2/m_H^2)$ from the region below $\sqrt{s} = m$ 
is negligibly small if $m_H^2 \gg m^2$. The dominant contribution 
comes from the asymptotic energy region.  When exponentiated, this 
integral generates an enhancement or suppression factor of 
$(m^*/m_H)^{2\delta(\infty)/\pi}$ for $M(s)$. If, for instance, 
the phase reaches $\delta(\infty)=\pm\pi/2$, the strong 
interaction correction would alter the amplitude by factor 16 
for $m_H = m_B$ and $m^* = m_W$. It means an enhancement or 
a suppression of factor $\sim 250$ in rate. There is no evidence 
for such a huge dynamical enhancement or suppression when we compare 
the observed two-body decay rates with the theoretical estimates
in the $D$ and $B$ decays. The so-called color suppression
observed in the B decay should be attributed not to a severe 
dynamical suppression by strong interactions but to lack of 
strong-interaction corrections, since the suppression exists 
prior to final-state interactions.   

The obvious  alternative to the asymptotic behavior of Eq.(\ref{asym}) is
\begin{equation}
        \delta(s) \rightarrow 0\;\;\;(m\leq\sqrt{s}\leq m_W).
\end{equation}
The enhancement or suppression is milder in this case.

The fractional power $(m^*/m_H)^{2\delta(\infty)/\pi}$ does not appear in 
the conventional calculation of the decay matrix elements. The short-distance 
QCD corrections enter in fractional powers of $(\ln m^*/\ln m_H)$.
{\it If} we wish to reproduce the short-distance correction factor of the
renormalization group with our final-state interaction integral, we 
would choose such that $\delta(s)$ approach zero asymptotically as
\begin{equation}
   \delta(s) \rightarrow \overline{\delta}/\ln s, 
                   \;\;\;(m\leq \sqrt{s}\leq m_W) \label{RGbehavior}
\end{equation}
where $\overline{\delta} = \gamma_w/b_0$, $\gamma_w$ is a constant 
determined by the anomalous dimension of $H_w$, and $b_0$ is 
from the running QCD coupling. The asymptotic behavior of 
Eq.(\ref{RGbehavior}) leads to $(\ln m^*/\ln m_H)^{2\overline{\delta}/\pi}$ 
for the amplitude. Since our correction factor includes 
both short- and long-distance effects, Eq.(\ref{RGbehavior}) 
should represent only the portion of the asymptotic phase 
that is attributed to the short-distance interactions. 
We must study the long-distance effects by some other means.

The main result of our dispersion relation is that the phase of the 
two-body decay amplitude of a heavy hadron should be zero or a small 
value for a large initial mass ($s\rightarrow\infty$). 
It should certainly not be $\pm 90^{\circ}$. It is worth noting  
that, unlike the phase, the magnitude of the amplitude can be subject
to substantial final-state interaction corrections since it picks up
the effects of all energies.

\section{Eigenphase shifts}

  In this Section we study the decay phases from the viewpoint 
of the eigenphases of the S-matrix of high-energy scattering. We need
to know about the composition of the eigenchannels and their high-energy
asymptotic behavior.  We shall use the dual resonance model as our guide 
since it is the most successful model that incorporates the relevant 
aspects of the quark model spectroscopy and long-distance hadron scattering.

The partial-wave S-matrix of strong-interaction is diagonalized in terms of
the eigenchannels carrying the quantum numbers of hadron $H$. Labeling the
eigenchannels by $|a\rangle, |b\rangle, \cdots$ we can express the S-matrix 
elements as
\begin{eqnarray}
        S_{ab} &=&  \langle b^{out}|a^{in}\rangle \\ \nonumber
               &=&  \delta_{ab} e^{2i\delta_a(s)}.
\end{eqnarray}
Without loss of generality we normalize the eigenphase shifts to zero
at their respective thresholds. Experimentally observed are the hadronic
states with each particle carrying definite momentum.  We project those 
states onto the $J^P$ eigenstates and denote them by $h$. By completeness 
of eigenchannels, $h$ can be expanded as
\begin{equation}
          |h\rangle = \sum_a O_{ha}|a\rangle, \label{expansion}
\end{equation}
where we can choose $O_{ha}$ to be an orthogonal matrix as a consequence of
T-invariance for strong interactions. 

The partial-wave amplitude $a^J(s)$ for elastic scattering is expanded 
in the eigenphase shifts as
\begin{equation}
        a^J(s) = \sum_a O_{ha}^2 e^{i\delta_a}\sin\delta_a, \label{scattering0}
\end{equation}
or in the real and imaginary parts,
\begin{eqnarray}
   {\rm Re}a^J(s) &=& \sum_aO_{ha}^2\cos\delta_a\sin\delta_a, \\ \nonumber
   {\rm Im}a^J(s) &=& \sum_aO_{ha}^2\sin^2\delta_a. \label{scattering}
\end{eqnarray}
Both $O_{ha}$ and $\delta_a$ are $s$-dependent.
Strictly speaking, once multibody states are included, we must label 
the eigenchannels with continuous parameters.  Therefore the discrete 
summation in Eqs.(\ref{expansion}) and (\ref{scattering}) is symbolic. 
The state density per unit energy in a volume characteristic of strong 
interactions ($\sim m_{\pi}^{-3}$) may substitute as the effective 
number of states for multibody channels.  

Actually the dual resonance model\cite{Veneziano} of the late 1960's 
answers to how many states exist at energy $\sqrt{s}$. In this model, 
the number of states was counted to be\cite{FVBM}
\begin{equation}
          n_0\sim \frac{1}{(\alpha's)^{(d+1)/2}}\exp(\sqrt{s}/m_0),
                      \label{statedensity}
\end{equation}
where $\alpha'$ is the Regge slope ($\simeq$1GeV$^{-2}$), $d$ is the 
space-time dimension, and $m_0= (3/2\alpha'd)^{1/2}/\pi$, 
which takes a value $\sim 200$MeV for $d = 4$. This state density $n_0$ 
contains the states of all angular momentum $J \leq (\alpha's)^{1/2}$ 
at mass $\sqrt{s}$ for the interval of $\sim$ 1GeV$^{-2} (\approx\alpha')$. 
While the highest $J$ state has no degeneracy, degeneracy of states rapidly 
increases with descending $J$. It also includes the states of negative 
norm on the daughter Regge trajectories in the case of 4-dimensional 
space-time.  Nonetheless the state density of Eq.(\ref{statedensity}),
particularly the exponential dependence, gives an order-of-magnitude 
estimate for the number of the J=0 states in the dual resonance model. 
Hagedorn\cite{Hagedorn} introduced a statistical model of hadrons
with quite a different motivation. It is amusing that his model 
led to essentially the same state density with a very close value 
($\sim$160 MeV) for $m_0$ but with a slightly difference power of 
$s$ in front. In Hagedorn's model, the closeness of $m_0$ to the pion 
mass was explained by the fact that every time energy increases 
by $m_0$, one more pion evaporates and causes the exponential growth 
of the state density.

The phenomenological success of the dual resonance model\cite{Lovelace} 
confirmed that the $s$-channel resonances are dual to the non-Pomeron 
Regge exchanges while the Pomeron is dual to the nonresonant continuum 
in the $s$-channel\cite{Collins}. The dual resonance model at tree-level
incorporates only the non-Pomeron trajectories. The Pomeron term was 
included by adding the two-body nonresonant intermediate states.
At high energies, the resonant states are so broad in width that 
they are not recognized as resonances but make up the smooth Regge 
asymptotic behavior of non-Pomeron exchange. In the yet higher 
energy region, the diffractive scattering dominates so that the s-channel 
states consist almost entirely of nonresonant states. Henceforth we 
shall call the $s$-channel states dual to the Pomeron and to 
the non-Pomerons as the nonresonant and the resonant channels, 
respectively, even though no resonance peak appears in the resonant 
channels at high energies. We shall also use the words, the diffractive 
and nondiffractive channels, for them. In the dual resonance model, 
the effective number of the nonresonant channels is even larger than 
$n_0$ of the resonant channels in the high-energy limit.
 
The distinction between the Pomeron and the non-Pomeron trajectories 
is best described by the quark diagram\cite{HR}. (See Figure 1.)  The 
non-Pomeron exchange in the boson-boson scattering is described by a pair 
of quark-antiquark in the intermediate state (Fig.1a) which represents 
a tower of resonances in the s-channel and at the same time the Regge 
trajectory exchanges in the $t$-channel. In contrast, the Pomeron 
exchange corresponds to the "disconnected" quark diagram of the 
$q\overline{q}q\overline{q}$ four-quark intermediate states in the 
$s$-channel as shown in Fig.1b. In the context of QCD one pair of 
$q\overline{q}$ exchanges relatively soft gluons with the other pair 
of $q\overline{q}$ in the Pomeron exchange\cite{LN}. This quark diagram 
not only ensures the exchange degeneracy for a pair of non-Pomeron 
trajectories with opposite signatures, as observed in scattering 
experiment\cite{Collins}, but also explains the absence of the mesons 
with exotic quantum numbers such as $I\geq 3/2$ in meson spectroscopy.

\subsection{Final-state interactions in eigenphase shifts}

The T-invariance relation in Section II can be easily extended 
to the decay amplitude into the eigenchannels. Using the 
completeness condition $\sum_a|a^{out}\rangle\langle a^{out}|=1$
and $\langle a^{out}|a^{in}\rangle = e^{2i\delta_a}$, we obtain
\begin{equation}
  \langle a^{out}|H_w|H\rangle
    = e^{2i\delta_a(s)}\langle a^{out}|H_w|H\rangle^*,
\end{equation}
where $s=m_H^2$. Accordingly the decay amplitude $M_a(s)$ 
into the eigenchannel $a$ carries the eigenphase $\delta_a(s)$. 
Inserting the complete set of eigenchannels and using
the expansion of $|h\rangle$, we obtain
\begin{equation}
   \langle h^{out}|H_w|H\rangle = \sum_a O_{ha}\langle a^{out}|H_w|H\rangle.
\end{equation}
The decay amplitude $M_h(s)$ for $H\rightarrow h$ is now expressed
in the terms of the eigenchannel decay amplitudes $M_a(s)$:
\begin{equation}
     M_h(s) = \sum_a O_{ha}M_a(s).  \label{weakamp}
\end{equation}
Separating the phase $\delta_a(s)$ from $M_a(s)$ as
\begin{equation}
     M_a(s) \equiv \overline{M}_a(s)e^{i\delta_a(s)}, \label{weakamp'}
\end{equation}
we put Eq.(\ref{weakamp}) in the form
\begin{equation}
     M_h(s) = \sum_a O_{ha}\overline{M}_a(s)e^{i\delta_a(s)}. \label{weakamp"}
\end{equation}
Comparing Eq.(\ref{weakamp"}) with Eq.(\ref{scattering0}), we clearly
see that the net phase of $M_h(s)$ has little to do with that 
of $a^J(s)$. The phase of $M_h(s)$ agrees with that of $a^J(s)$,
barring an accident, only when there is only one eigenchannel
so that the elastic unitarity holds,
\begin{equation}
       |1+2ia^J(s)| = 1.  \label{unitarity}
\end{equation}
It was pointed out that the partial-wave projection of the diffraction 
amplitude for the $\pi\pi$ scattering at the $B$ mass is far short of 
the unitarity limit\cite{Donoghue}; $|1+2ia^J(s)| \simeq 0.6$.
For the $D$ decay, which occurs near the resonance region or a little 
above it, the nonleading Regge terms cannot be ignored in two-body 
scattering. We shall see in Section V that even after adding the 
non-Pomeron terms the elastic unitarity does not hold at the $D$ mass. 

The amplitude $\overline{M}_a(s)$ defined in Eq.(\ref{weakamp'}) 
still contains strong-interaction effects. To define the eigenchannel 
decay amplitude free of final-state interactions, we must separate 
out not only the phase correction but also the magnitude correction 
$\Delta_a(s)$ by $\overline{M}_a(s)=\overline{M}_{0a}(s)\Delta_a(s)$. 
If all inelastic channels are approximated as two-body or quasi-two-body
states, $\Delta_a(s)$ can be written in the Omn\`{e}s-Mushkelishvili 
representation:
\begin{equation}
    \Delta_a(s) = \exp\biggl(\frac{\cal{P}}{\pi}\int_{s_{0a}}^{m_W^2}
        \frac{\delta_a(s')}{s'-s}ds'\biggr),
\end{equation}
where $\cal{P}$ stands for the principal value integral. One obvious 
property of $\Delta_a(s)$ is that it is positive definite. Whether 
$\Delta_a(m_H^2)$ gives an enhancement ($>1$) or a suppression ($<1$) 
depends on the sign of $\delta_a(s)$ for all values of $s$ from 
the threshold $s_{0a}$ to $m_W^2$, not just the on-shell value 
$\delta(m_H^2)$. In the nonresonant channels, we shall argue later that 
$\delta_a$ is small in magnitude and the sign of $\delta_a$ can be easily 
flipped as energy changes when a large number of channels mix. 
If so, a correlation of the magnitude correction $\Delta_a(m_H^2)$ 
with the on-shell phase value $\delta_a(m_H^2)$ is tenuous, if any.  
With the magnitude enhancement factor written out, Eq.(\ref{weakamp"}) 
turns into
\begin{equation}
     M_h(s) = \sum_a O_{ha}\overline{M}_{0a}(s)\Delta_a(s)e^{i\delta_a(s)}.
                                                     \label{weakamp3}
\end{equation}

For the resonant channels, one can compute the Feynman diagram for the 
decay process $H\rightarrow R$(resonance) $\rightarrow h$(two-body) through 
a resonance $R$. The decay amplitude for the resonant eigenchannel 
$r$ takes the form, 
\begin{equation}
       M_r(s) = f_{HR}\frac{\sqrt{m_R\Gamma_r}}{m_R^2-s-im_R\Gamma_{tot}},
\end{equation}
where $f_{HR}$ is the $H$-$R$ pole transition strength and 
$\Gamma_{tot, r}$ are the total and partial decay widths of $R$. 
Note that $\Gamma_{tot}=O(m_H)$. In the quark model $f_{HR}$ is 
the overlap of the wave functions of $H$ and $R$. Since the overlap 
does not increase with $s$,  $M_r(s)$ decreases like $1/s$ or 
faster as $s\rightarrow\infty$.  If we express the resonant channel
contributions in the form of Eq.(\ref{weakamp3}), it means 
$\overline{M}_{0r}(s)\Delta_r(s)\rightarrow 1/s$ or faster
for each resonant channel $r$. 

\subsection{Strong interaction scattering}

Elastic scattering provides useful pieces of information 
about the eigenphase shifts. Experiment shows that the 
imaginary part dominates over the real part for the invariant 
amplitude $T(s,t)$ of elastic scattering at high energies.
Theoretically, the dominance of Im$T(s,t)$ is a general 
consequence of analyticity and crossing symmetry, not specific 
to the Regge theory, when the total cross section approaches 
a constant up to powers of $\log s$. The amplitude of the 
flat Pomeron trajectory,
\begin{equation}
        T(s,t) = i s\sigma_{tot}e^{bt}, \label{diffractionamp}
\end{equation}  
gives a reasonably good description of the diffractive scattering in the 
whole high-energy region relevant to us.\footnote{At energies above
$\sqrt{s} \approx 100$ GeV, the total cross sections actually rise very 
slowly with energy. One fit to $pp$-collisions gives $\sigma_{tot}(s)\simeq 
\beta_P(s/s_0)^{0.08}+\beta_{\rho-f}(s/s_0)^{-0.56}$\cite{Donnachie}.
This $s$-dependence requires that the forward scattering amplitude 
contains a real part:
     ${\rm Re}T(s,0)/{\rm Im}T(s,0) \rightarrow \tan(\pi\epsilon/2)$,
where $\epsilon = 0.08$. The forward scattering amplitude contains 
a real part by about 10\% even at extremely high energies. However 
such high energies have no direct relevance to the final-state
interaction of the B decay.} Eq.(\ref{diffractionamp}) leads 
to the elastic cross section,\footnote{We ignore all hadron masses as 
compared with $\sqrt{s}$ throughout this section.}
\begin{equation}
   \sigma_{el} = \sigma_{tot}^2/32\pi b,
\end{equation}
and to the partial-wave amplitude
\begin{equation}
      a^J(s) = i\sigma_{tot}/16\pi b,\;\;
       (J \ll (s/s_0)^{1/2}).   \label{partialwave}
\end{equation}
When we parametrize $T(s,t)$ by Eq.(\ref{diffractionamp}), the partial-wave
elasticity $\sigma_{el}^J/\sigma_{tot}^J\;(= {\rm Im}a^J(s))$ 
for $J \ll (s/s_0)^{1/2}$ is equal to twice the total elasticity, 
$(1/\sigma_{tot})\int(d\sigma_{el}/dt)dt$. The values for $\sigma_{tot}$ 
and $b$ can be extracted from the experimental data on $pp$, $\pi p$, 
and $Kp$ scattering.  The factorization of the Regge residues\footnote{
The factorization can be proved only for relatively simple J-plane 
singularlities. It is an assumption for the more general Pomeron.} 
relates the high-energy total cross sections by
\begin{equation}
    \sigma^{MM'}_{tot}
        = \sigma^{Mp}_{tot}\sigma^{M'p}_{tot}/\sigma^{pp}_{tot},
\end{equation}
where $M$ and $M'$ stand for mesons.
With $\sigma^{pp}_{tot} = 37$mb, $\sigma^{\pi p}_{tot} = 21$mb, and 
$\sigma^{Kp}_{tot}=17$mb for the diffractive contribution of 
$\sigma_{tot}$ at $\sqrt{s}=2\sim 8$ GeV\cite{Barger}, we obtain
\begin{equation}
    \sigma^{\pi\pi}_{tot} = 12{\rm mb}\;\;\;
    \sigma^{K\pi}_{tot} = 10{\rm mb}. 
\end{equation}
The numerical values are roughly in line with the empirical law of the 
quark number counting for the total cross sections: $\sigma^{MM'}_{tot} :
\sigma^{Mp}_{tot}: \sigma^{pp}_{tot} = 2^2 : 2\times 3 : 3^2$.
The diffraction width parameter $b$ obeys the inequality $b_{pp} >
b_{\pi p} > b_{Kp}$.  In one analysis\cite{Chow}
\begin{equation}
      b_{pp}\simeq 5{\rm GeV}^{-2}, \;\;\;
      b_{\pi p}\simeq 4.3{\rm GeV}^{-2},\;\;\;
      b_{Kp}\simeq 3.2{\rm GeV}^{-2}.
\end{equation}
The parameter $b$ is related to the effective target size of colliding 
hadrons in elastic scattering. The inequality $b_{pp} > b_{\pi p}$ 
indicates that the proton is a little more spread than the pion.  
The electromagnetic form factors show the same trend: 
$(r_p^2/6)^{1/2} \simeq \sqrt{2}m_{\rho}^{-1}$ and
$(r_{\pi}^2/6)^{1/2} \simeq m_{\rho}^{-1}$. The relation $b_{Kp} < 
b_{\pi p}$ may be interpreted as a result of the less intense interaction 
of the s-quark with the u/d-quarks, which reduces the effective size of 
$K$ as well as $\sigma^{Kp}_{tot}$. This line of argument leads us to the 
$b$-parameters for $\pi\pi$ and $K\pi$ scattering somewhere around
\begin{equation}
    b_{\pi\pi}\simeq 3.7{\rm GeV}^{-2}, 
    \;\;\; b_{K\pi}\simeq 2.8{\rm GeV}^{-2}.
\end{equation}
If we extrapolate this reasoning to the $D\pi$ scattering and ignore the 
c-quark interaction with the u/d-quarks, we are led with $\sigma^{D\pi}_{tot}
= \sigma^{\pi\pi}_{tot}/2$ and $b_{D\pi}<b_{K\pi}$ to
\begin{equation}
     \sigma^{D\pi}_{tot} = 6{\rm mb}\;\;\;
     b_{D\pi}  \simeq 2.4{\rm GeV}^{-2}.  
\end{equation} 
When we substitute these values of parameters in the partial-wave 
projection of Eq.(\ref{partialwave}), we obtain 
\begin{equation}
        {\rm Im}a^J(s) \simeq \left\{ \begin{array}{cc}
                            0.16 &(\pi\pi) \\
                            0.17 &({\rm K}\pi) \\
                            0.12 &({\rm D}\pi)  \end{array} \right.
                            \label{estimate}
\end{equation}
The value of Im$a^J(s)$ is even smaller for $\psi\pi$ and $\psi K$.
The precise values of the right-hand side of Eq.(\ref{estimate}) are not
important in the following. It is no surprise that the elastic unitarity 
is not satisfied for any of the above processes:
\begin{equation}
      |1+2ia^J(s)| = 0.66\sim 0.76.
\end{equation}
Note that $|1+2ia^J(s)|=0.5$ is the limit of the completely absorptive 
black target. When $a^J(s)$ is purely imaginary, the value of $|1+2ia^J(s)|$ 
does not tell the whole story.  The partial-wave inelasticity reveals more of 
the scattering mechanism.  The numbers in Eq.(\ref{estimate}) give
the partial-wave inelasticity (=1$-$Im$a^J$) in the range of
\begin{equation}
      \sigma_{inel}^J/\sigma_{tot}^J = 0.83\sim 0.88.
\end{equation} 
 
It should be pointed out here that the $s$-wave phase shift is dominated by 
long-distance physics.  The largest contribution to $a^J(s)$ for small $J$ 
comes from the region of the momentum transfer $-1/b <t \leq 0$.
The contribution from the perturbative QCD region of large $|t|$ is
small. It is $\sim 1/|t|^{1/2}$ that determines the effective 
distance of interactions. It is true, however, that the $s$-wave contains 
a larger share of short-distance physics than high partial-waves. 

Comparing Eq.(\ref{estimate}) with Im$a^J$ of Eq.(\ref{scattering0}), 
we find that the average or typical eigenphase shift of the diffractive 
channels should be in the range of 
\begin{equation}
                 \sin^2\delta_d \simeq 1/8 \sim 1/6, \label{ave}
\end{equation}
where the subscript $d$ for the phase shift stands for "diffractive".
The smallness of Re$a^J$/Im$a^J$ for the diffractive scattering
suggests that $\delta_d\;(d=1,2\cdots)$ spreads over positive 
and negative values (modulo $n\pi$) in an approximately 
symmetric distribution with respect to $\delta_d = 0$. 
For such a distribution a large cancellation occurs among different
eigenchannels in Re$a^J(\propto\cos\delta_d\sin\delta_d)$ while every 
term adds up in Im$a^J(\propto\sin^2\delta_d)$. What about the $n\pi$ 
ambiguity for $\delta_d$?  Since the diffractive channels are nonresonant, 
$\delta_d$ starts at 0 and does not go over $\pi/2$. Because of Wigner's 
theorem\cite{Newton} on the causality constraint on phase shifts, it is 
not very likely for $\delta_d$ to turn clockwise and go over $-\pi/2$.
Therefore $\delta_d$'s stay in $-\pi/2<\delta_d<\pi/2$, spreading 
symmetrically with respect to $\delta_d=0$ over the range roughly
\begin{equation}
         - \pi/8 < \delta_d < \pi/8.
      \label{range}
\end{equation}    
Here again, the precise values of the upper and lower bounds 
are unimportant to us.

Even in the decay into a diffractive multibody channel a resonance can 
appear in a subchannel giving rise to a large phase when the subchannel 
invariant mass coincides with the resonance mass. For instance, 
$H\rightarrow R\pi$(nonresonant)$\rightarrow K\overline{K}\pi$ at
$m_{K\overline{K}}=m_R$. If $R$ is a sharp resonance of $K\overline{K}$, 
we treat the process as a two-body decay into $R\pi$.  If not, the 
final-state is a three-body channel. Though the decay amplitude has a 
large phase at $m_{K\overline{K}} = m_R$, this phase is washed out after 
the $m_{K\overline{K}}$ is integrated over with $\sqrt{s}$ fixed to $m_H$.

Let us turn to the nondiffractive channels. Their contribution to
$\sigma_{tot}$ falls off like $1/s^{1/2}$ or faster relative 
to that of the diffractive channels. To obtain the subdominant terms 
in $s$, we add the nonleading Regge contributions.  One may wonder 
about a possible isolated $s$-channel non-Regge singularity which may
contribute only to a single angular momentum. Especially relevant is 
the fixed singularity $a^J \propto\delta_{J0}/J$. This singularity
generates a constant term in $T(s,t)$ for all values of $s$ and $t$.
Since hadrons are composites of the scale $\Lambda_{QCD}$, there must 
not be such a hard interaction.

According to the Regge duality, the nondiffractive and diffractive portions 
of the amplitude are separately dual to the resonant and nonresonant states,
respectively, in the $s$-channel.  Therefore the nondiffractive or resonant 
contribution $a_r^J(s)$ to the partial-wave amplitude can be expressed as
\begin{equation}
      a_r^J(s) = \sum_r O_{hr}^2 e^{i\delta_r}\sin\delta_r, 
\end{equation}
where the subscript $r$ stands for "resonant".
In these channels, the eigenphase shift $\delta_r$ turns around counter 
clockwise slowly passing $\pi/2$ at the resonance and approaches $\pi$
asymptotically. Comparison of the asymptotic energy dependences of the 
diffractive and nondiffractive amplitudes gives us
\begin{equation}
      \frac{\sum_r O_{hr}^2\sin^2\delta_r(s)}{\sum_d O_{hd}^2\sin^2\delta_d(s)}
  = (\overline{\beta}_{\rho{\rm -}f}/\overline{\beta}_P)s^{-1/2}, 
      \label{nonleading}
\end{equation}
where $\overline{\beta}_{P, \rho{\rm -}f}$ are the properly normalized 
Regge residues at $t=0$ of the Pomeron and the $\rho$-$f_2$ trajectory, 
and $s$ is in the unit of 1 GeV$^2$.  We can extract the Regge residues 
from the energy dependence of total cross sections. For $\pi^+p$ $(\pi^-p)$ 
scattering\cite{Barger}, for instance,
\begin{equation}
 \overline{\beta}_{\rho{\rm -}f}/\overline{\beta}_P\simeq 0.75 (1.18). 
     \label{nonleading'}
\end{equation}
We can relate the left-hand side of Eq.(\ref{nonleading}) to the numbers of
the resonant and nonresonant eigenchannels. By replacing $\sin^2\delta_r$ 
by unity for the resonant channels and substituting the average value
for $\sin^2\delta_d$, we obtain
\begin{equation}
    \frac{\sum_rO_{hr}^2\sin^2\delta_r(s)}{\sum_dO_{hd}\sin^2\delta_d(s)}
     \simeq \frac{n_r}{(n - n_r)\langle\sin^2\delta_d\rangle}.
\end{equation}
With Eqs.(\ref{nonleading}) and (\ref{nonleading'}), we obatin
\begin{equation}
   n_r \approx n\langle\sin^2\delta_d\rangle/s^{1/2},
\end{equation}
which agrees with our expectation $n_r \ll n$.

\section{Random S-matrix approximation}

When a very large number of eigenchannels are present, studying individual
channels is impractical.  To study the spacings and widths for hundreds of 
the densely populated resonances in complex nuclei, nuclear physicists 
introduced a certain randomness hypothesis in the multichannel S-matrix. 
The work was started by Wigner\cite{Wigner}, pursued by many\cite{Many},
and brought into a mathematical sophistication by Dyson\cite{Dyson}. 
It succeeded in reproducing various features of those resonances\cite{Many}.
A similarity of the nuclear resonances to the multitude of the hadron 
channels in our problem suggests us to study physics of the uncontrollably 
many eigenchannels with the randomness hypothesis.

The randomness of the channel mixing $O_{ha}$ can arise in our problem
when a very large number of eigenchannels exist in degeneracy with 
the two-body final-states. Since even a small coupling strongly mixes 
a pair of degenerate states, any two-body state in the highly inelastic 
energy region is a linear combination of many eigenchannels. The expansion 
coefficients of a given state into eigenchannels, namely $O_{ha}$, is 
sensitive to the strength of channel couplings, but many features of 
physics, for instance $\delta_a$, should not be sensitive to small 
variations of channel couplings. It is therefore reasonable to postulate 
that quantities of our interest can be computed by replacing products 
of $O_{ha}$ with their statistical averages over the phase space of 
the $O(n)$ rotations of $O_{ha}$. After the $O(n)$ average has been taken
for products of $O_{ha}$, we are left with $\delta_a$. In our problem 
unlike the nuclear resonances, we cannot postulate a complete randomness 
for the distribution of $\delta_a$. If we assumed that both $\delta_a$ and 
$O_{ha}$ are completely random or chaotic, Eq.(\ref{scattering}) would 
lead us to $\langle O_{ha}O_{kb}\rangle = \delta_{ab}\delta_{hk}/n$,
$\langle\sin^2\delta_a\rangle = 1/2$, and
$\langle\cos\delta_a\sin\delta_a\rangle = 0$. Consequently  
\begin{equation}
      {\rm Re}a^J = 0, \;\; {\rm Im}a^J = 0.5.
\end{equation}
This corresponds to the scattering from a black disc that gives the 50\%
elasticity due to the shadow scattering in disagreement with 
Eq.(\ref{estimate}). In the actual high-energy scattering,
$\sigma_{el}/\sigma_{tot}$ is considerably less than 50\%.  It means 
that a hadron target behaves like an opaque disc at high energies.
To describe such a target, we must postulate randomness only within 
the restricted range as specified in Eq.(\ref{range}).\footnote{We 
previously studied mainly the magnitude of the squared decay amplitude 
in this statistical model\cite{CS}. Here we focus on the decay phase by 
refining some of the previous postulates. In Ref\cite{CS}, the magnitude 
factor $\Delta_a(s)$ was not separated but ignored for simplicity.} 

\subsection{Dominance of the real part for decay amplitudes}

We start with Eq.(\ref{weakamp3}) and isolate all eigenchannel 
dependences by expressing $M_a(s)=\overline{M}_{0a}\Delta_a 
e^{i\delta_a}$ back in terms of the decay amplitudes of the 
hadron basis $k$ which are free of the final-state interaction 
(of $\sqrt{s}\leq m_W$). We denote such decay amplitudes 
by $\overline{M}_{0k}(s)$. Writing the nonresonant 
and resonant channels separately, we have
\begin{equation}
      \overline{M}_{0k}(s) = \sum_d O_{kd}\overline{M}_{0d}(s)
                        + \sum_r O_{kr}\overline{M}_{0r}(s).
\end{equation}
Substituting the inverted relation of this in Eq.(\ref{weakamp3}), we obtain
\begin{equation}
   M_h(s) = 
    \sum_{d,k}O_{hd}O_{kd}\overline{M}_{0k}(s)\Delta_d(s)e^{i\delta_d(s)}
    +\sum_{r,k}O_{hr}O_{kr}\overline{M}_{0k}(s)\Delta_r(s)e^{i\delta_r(s)}.
                   \label{randomM}
\end{equation} 
Using $\langle O_{ha}O_{kb} \rangle = \delta_{hk}\delta_{ab}/n$, we obtain 
the random phase values for the real and imaginary parts of $M_h(s)$:
\begin{equation}
    M_h(s) = \overline{M}_{0h}(s)(\langle\Delta\cos\delta\rangle
   + i\langle\Delta\sin\delta\rangle),  \label{averageM}
\end{equation}
where
\begin{equation}
  \langle\Delta\cos\delta\rangle=\frac{1}{n}\biggl(
 \sum_d\Delta_d(s)\cos\delta_d(s)+\sum_r\Delta_r(s)\cos\delta_r(s)\biggr),
   \label{averageDelta}
\end{equation}
and $\cos\delta\rightarrow \sin\delta$ for 
$\langle\Delta\sin\delta\rangle$.

Let us leave out the resonant eigenchannels for the moment.  Then, first 
of all, the phase of $M_h(s)$ approaches a common limit in Eq.(\ref{averageM})
independent of isospins or charge states of $h$, say, $K^-\pi^+$ and 
$\overline{K}^0\pi^0$. The common limit does not depend on the effective 
weak interaction $H_w$.  This marked simplicity is valid only in the random
phase limit. Considering that Im$a^J$ takes roughly the same 
value for all meson-meson scatterings (cf. Eq.(\ref{estimate})), we expect
the common decay phase is not very sensitive to the final hadrons. 

We have pointed out earlier that in the quasi-two-body approximation to 
the inelastic channels, $\Delta_a(s)$ is positive definite and uncorrelated 
with the on-shell $\delta_a(s)$. If this is the case, the terms of different
eigenchannels cancel each other in Im$M_h(s)$ because of the random signs of
$\sin\delta_d$:\footnote{The possibility of many phases averaging out to 
a small net decay phase was mentioned earlier by Wolfenstein\cite{Wolfenstein} 
to the author.} 
\begin{equation}
  {\rm Im}M_h(s)= \sum_{\delta_d>0}\Delta_d(s)\sin\delta_d
      - \sum_{\delta_{d'}<0}\Delta_{d'}(s)|\sin\delta_{d'}|. \label{sum}
\end{equation} 
In contrast, all eigenchannels add up in the real part since $\cos\delta_d>0$ 
in the restricted range of Eq.(\ref{range}). It is similar to the situation 
in the elastic scattering amplitude of Eq.(\ref{scattering}), but this time 
the cancellation occurs in the imaginary part instead of the real part. 
Because of the deviation of $\Delta_d(s)$ from unity due to enhancement 
and suppression, the cancellation may not be as good as in the scattering.
We can set with Eqs.(\ref{averageM}) and (\ref{averageDelta}) a loose
upper bound on the imaginary-to-real ratio for $M_h(s)$. In terms of
$\tan\delta_h = {\rm Im}M_h(s)/{\rm Re}M_h(s)$, it is given by\footnote{In 
Section II, $\delta_h$ was simply written as $\delta(s)$.}
\begin{equation}
  |\tan\delta_h|  < |\langle\sin\delta\rangle|/\langle\cos\delta\rangle.
                                           \label{ImM}
\end{equation}
The right-hand side is less than $\sim$0.4 acccording to Eq.(\ref{ave}).
This number would be realized when no cancellation occurs in Eq.(\ref{sum}). 
Since $\Delta_d(s)$ and the sign of the on-shell $\delta_d(s)$ 
are only tenuously correlated, we expect in reality a fairly high 
degree of cancellation between the terms of $\delta_d>0$ and 
$\delta_d<0$. Therefore the actual value of the imaginary-to-real 
ratio is most likely much smaller than 0.4. By pushing our 
approximation further, let us set $\Delta_d$ to a common number
\footnote{If the sign of the on-shell $\delta_d$ is not correlated 
with $\Delta_d>1$ or $<1$, setting $\Delta_d$ to a number independent
is justifiable.} and $\cos\delta_d \simeq 1$.  Then we obtain
\begin{equation}
     \tan\delta_h \simeq \frac{{\rm Re}a^J}{{\rm Im}a^J}
                     \langle\sin^2\delta\rangle. \label{deltaestimate}
\end{equation} 
The right-hand side is zero for the purely diffractive amplitude. Even if 
we use the parametrization $\sigma_{tot}\sim s^{0.08}$ fitted to much 
higher energies which requires a nonnegligible real part for $T(s,0)$,
the right-hand side is $\simeq 0.04\pi\times\langle\sin^2\delta\rangle\simeq
2\times 10^{-2}$. 

  Next we look into the contribution of the resonant channels to the 
decay phase. It falls like $s^{\alpha_{\rho{\rm -}f}-1}\simeq s^{-1/2}$ 
relative to the diffractive contribution. In addition to
the $s^{-1/2}$ suppression, the chiral structure of the weak
interaction suppresses the transition from the initial state $H$ of
$J^P = 0^-$ to the resonant channels. Let us examine this suppression.

   Keeping in mind that in the dual resonance model the resonant
channels are made of two-quark states in the quark diagram (Fig.1a),
we examine the transition of $H (= \overline{q}Q)$ to the resonant
state $R (= \overline{q}q')$.\footnote{The following argument is valid
also for $H\rightarrow R(=\overline{q}''q')$ with minor modifications.}
The bare weak Hamiltonian which causes the transition is the four-quark
operator,
\begin{equation}
 H_w = 4(G/\sqrt{2})V^*_{qq'}V_{qQ}(\overline{q'}_Lq_L)(\overline{q}_LQ_L),
\end{equation}
where we have suppressed the Dirac structure of the quark bilinears.
The short-distance QCD corrections to $H_w$
induce the effective interactions such
as $(\overline{q'}_L(\lambda_a/2)q_L)(\overline{q}_L(\lambda_a/2)Q_L)$
and the penguin interaction $(\overline{q}_R(\lambda_a/2)q_R)
(\overline{q}'_L(\lambda_a/2)Q_L)$. The weak transition $H\rightarrow R$
is the so-called annihilation or exchange process. The matrix elements of
the annihilation and exchange processes are suppressed by the decay constant
$f_H (=O(f_{\pi}))$ of the meson $H$ and, for the non-penguin interactions,
also by the factor $(m_{q'}+m_q)$ of chirality mismatch on the side of $R$.
The suppression due to the first one, $f_H/m_H$, together with
the energy dependence suppression $s^{-1/2}$ or $n_r/n$ is severe
enough to make the phase difference contribution of the resonant
channels negligibly small. It should be reminded that the {\it resonances}
in the resonant channels are so broad ($\Gamma_H = O(m_H)$) that
no sharp resonance enhancement arises.
The suppression of $f_H/m_H$ occurs for the transition
$H\rightarrow R$ through the penguin interaction too.

In short, the dominant decay process is through the spectator diagrams which
lead to $q\overline{q}q\overline{q}$ not $q\overline{q}$ in the final state.
The decay $H\rightarrow q\overline{q}q\overline{q}$ followed by a pair
annihilation into $q\overline{q}$ is no other than the annihilation
or exchange process, as we can see by drawing the diagram. The rescattering
of $q\overline{q}q\overline{q}$ without a $q\overline{q}$ annihilation is a
diffactive process not a resonant one in the sense of the dual resonance
model. Therefore, we can conclude that the decay phase due to the
resonant channels are negligibly small. We estimate a typical magnitude
of the phase differences of this origin as
\begin{equation}
     \Delta\delta \approx (f_H/m_H)(1{\rm GeV}/s^{1/2})
                       \label{phasedifference}
\end{equation}
for the decay modes where the spectator process dominates.
When the chiral mismatch occurs for $R$, a decay matrix element
is suppressed by the additional factor of $m_{q'}/m_H$.

\subsection{Phase difference between amplitudes}

What we can measure in experiment is not the absolute 
phases but the phase differences. Though the phase of 
$\langle M_h(s)\rangle$ is independent of the charge or isospin 
states of $h$, the fluctuation around the average value can be isospin 
dependent and therefore generate phase differences. The physical 
origin of this type of phase differences can be explained as 
follows: A pair of decay amplitudes have different compositions of
diffractive eigenchannels.  Therefore the interference between 
eigenchannels sums up to different net phases for the amplitudes.
The phases of this origin are washed out in the random limit. 
The interference between the diffractive and nondiffractive
eigenchannels is another possible source of the relative phases. 

We first examine the relative phase due to the fluctuations.
The fluctuation is the standard deviation from the random phase limit 
of $\langle M_h(s)\rangle$. Averaging out the product of four $O_{ha}$ 
in the phase space of O(n) group\footnote{It is straightforward to obtain
$\langle O_{ka}O_{la}O_{mb}O_{nb}\rangle = (\delta_{kl}\delta_{mn}+
\delta_{km}\delta_{ln} + \delta_{kn}\delta_{lm})/n(n+2)$ for $a=b$ and
$=[(n+1)\delta_{kl}\delta_{mn} - \delta_{kn}\delta_{lm} - \delta_{kn}
\delta_{lm}]/(n-1)n(n+2)$ for $a\neq b$\cite{CS}.}, we obtain for $n \gg n_r$
\begin{eqnarray}
  |\Delta{\rm Im} M_h(s)|^2 &\equiv&\langle({\rm Im}M_h(s)-
                  \langle{\rm Im} M_h(s)\rangle)^2\rangle,     \\ \nonumber
            &=& \langle |\overline{M}_0|^2\rangle 
                (\langle\Delta^2\sin^2\delta\rangle
               - \langle\Delta\sin\delta)\rangle^2)
               + O\biggl(\frac{1}{n}|M_h(s)|^2\biggr), \\ \nonumber
  |\Delta{\rm Re} M_h(s)|^2 &=& (\sin\delta \rightarrow \cos\delta),
              \label{SD}
\end{eqnarray}  
where the brackets denote the averages such as
\begin{equation}
  \langle|\overline{M}_0|^2\rangle = \frac{1}{n}\sum_d|\overline{M}_{0d}|^2.
\end{equation}
The sum $\sum_{a=(d,r)}|\overline{M}_{0a}|^2/2m_H$ is 
the total rate for the decays induced by $H_w$ without 
strong interaction corrections. Since the decay amplitude 
is dominantly real, the phase difference is more sensitive 
to $\Delta{\rm Im}M_h(s)$ than to $\Delta{\rm Re}M_h(s)$. 
The relative magnitude of the standard deviation $\Delta M_h(s)$ 
to the average $|M_h(s)|$ depends on the channel number $n$. 
If $n$ is nearly as large as the dual resonance model 
indicates, or even a small fraction of it, $|\Delta M_h(s)|$ is 
negligibly small and no relative phase arises from fluctuations. 
In this case, the two-body channel couples to an enormous number of 
inelastic nonresonant channels though its couplings to individual 
channels are very weak accordingly. The fluctuations almost even out when 
this happens. On the other hand, if $n$ is so small that $1/n$ 
is comparable to the branching fraction to the channel $h$, 
$\Delta{\rm Im}M_h(s)$ can be close to $|{\rm Im}M_h(s)|$ itself.
When $|M_h(s)|$ is small by an accidental cancellation of high
degree among different eigenchannels, the fluctuation can be 
large in proportion. To ensure that the random phase is a good 
approximation, therefore, we should apply its predictions only 
to the two-body modes having relatively large branching fractions.

The phase differences arising from an interference with the resonant 
eigenchannels are suppressed by the chiral structure of the weak interaction
operators and decrease like $s^{-1/2}$ with energy. Their contribution
is of the order of $\Delta\delta$ as given in Eq.(\ref{phasedifference}) in 
general. However, our argument fails in the modes where the spectator 
process is forbidden or highly suppressed. In such processes the diffractive 
eigenchannels dual to the Pomeron are missing or highly suppressed
so that large phases and phase differences may potentially arise.
It is unfortunate that the decay branching is generally very small 
for them. 

\section{The D and B decays}

We shall look into the specific cases of the D and B mesons with
a few critical remarks on some of the recent attempts to compute 
the final-state interaction phases of the two-body decays of $D$ and $B$.

\subsection{The D meson}

The most prominent two-body decay of the D meson is the $\overline{K}\pi$
modes. From the observed decay rates of $D^+\rightarrow
\overline{K}^0\pi^+$ and $D^0\rightarrow\pi^+K^-, \overline{K}^0\pi^0$, 
the phase and magnitude of the decay amplitude were determined for 
$D\rightarrow \pi\overline{K}$ of definite isospin.  They discovered
a large relative phase between the I=1/2 and 3/2 decay amplitudes\cite{Bishai};
\begin{equation}
         \delta_{3/2}(m_D^2) -\delta_{1/2}(m_D^2) = (96\pm 13)^{\circ} 
                \label{phase}
\end{equation}
with the amplitude ratio
\begin{equation}
         |M_{3/2}/M_{1/2}| = 0.27 \pm 0.03.
\end{equation}
Since the D meson mass is too low for the Pomeron to dominate in the 
$K\pi$ scattering, our argument in the preceding Section does not 
apply with a good accuracy. The phases of the $s$-wave $K\pi$ elastic 
scattering amplitudes $a_I(s)$ at energy $m_D$ was computed
with the Regge theory and identified  with the phases of the decay 
amplitudes for the $K\pi$ isospin eigenstates. 

The result of calculation in Ref.\cite{Weyers} happens to be in a good 
agreement with the observed phase difference, Eq.(\ref{phase}), within 
the uncertainties in the values of the Regge parameters.  
We argue however that such an agreement is fortuitous. In order to identify 
the scattering phases with the decay phases, the scattering  must 
be elastic.  The authors of Ref\cite{Weyers} computed only the phase 
difference of the $K\pi$ amplitudes, not the magnitudes of them. 
If they had done so, they would have obtained
\begin{eqnarray}
              a_{1/2}(m_D^2) &=& 0.08 +0.21i \\ \nonumber
              a_{3/2}(m_D^2) &=& -0.35 + 0.18i.  \label{unitary}
\end{eqnarray}
They do not satisfy the elastic unitarity: $|1+2ia_{1/2}| = 0.60$ and
$|1+2ia_{3/2}| = 0.95$. The $I=1/2$ amplitude is deep inside the Argand 
diagram while the $I=3/2$ amplitude nearly satisfies the elastic unitarity.  
Actually, if one insists that $K\pi$ scattering be elastic at $\sqrt{s} = 
m_D$, one can make $a_{3/2}$ unitary in a model-independent way. Since 
the $I=3/2$ channel is exotic, {\it i.e.,} contains no $s$-channel resonances 
dual to the non-Pomeron trajectories, the imaginary part of $a_{3/2}$ in 
Eq.(\ref{unitary}) comes entirely from the Pomeron. The non-Pomeron 
trajectories can contribute only to the real part in the case of $a_{3/2}$. 
Therefore one should make up for the unitarity violation of $a_{3/2}$ by 
adjusting its real part:
\begin{equation}
       a_{3/2}(m_D^2) = -0.38 + 0.18i, \label{unitarity1}
\end{equation}
which gives
\begin{equation}
       \delta_{3/2}(m_D^2) = 155^{\circ}.
\end{equation}
Unlike $a_{3/2}$ the unitarity deficit of $a_{1/2}$ is very large. 
To recover the elastic unitarity for $a_{1/2}$, the Pomeron daughters and 
the exchange-degenerate pairs of trajectories such as the daughters of 
$\rho$-$f_2$ and of $K^*$-$K_2$ must contribute just as much as or more 
than the parent trajectories. If so, the Regge expansion itself would be 
unjustified at this energy.  
 
In the $I=1/2$ channel there is an obvious candidate for a cause of 
inelasticity, that is, $\overline{K}\eta$ channel which should 
enter according to the flavor SU(3) symmetry. Then the 
$(\overline{K}\pi)_{I=1/2}$ involves at least two eigenchannels, 
the $I=1/2$ channels of the {\bf 8} and {\bf 27} representations of 
SU(3). In addition, there are multibody channels. The $D$ meson decays into 
$\overline{K}\pi\pi\pi$ of both $I=1/2$ and $3/2$ including the modes such as
$\overline{K}^*\rho$ and $\overline{K}a_1$. The combined branching fraction
of all $\overline{K}\pi\pi\pi$ modes is even larger than that of 
$\overline{K}\pi$. Many $J^P=0^+$ states are contained in 
$\overline{K}\pi\pi\pi$, with which $\overline{K}\pi$ mixes in general. 
It is not surprising if many eigenchannels are open for both $I=1/2$ 
and $3/2$ already at the energy of the $D$ meson

Is there any chance to identify the $D\rightarrow\overline{K}\pi$ decay 
phase with the $\overline{K}\pi$ scattering phase all the way up 
to high energies despite the presence of inelastic channels ? 
We do not think so. In Eq.(\ref{unitary}) the real part of $a_{I}$ 
is positive for $I=1/2$ and negative for $I=3/2$ at the $D$ meson mass.  
If it happens that $a_{1/2}$ and $a_{3/2}$ approach purely imaginary 
numbers at high energies, $\delta_{1/2}(s)$ would turn counter clockwise 
to $\pi/2$ and $\delta_{3/2}(s)$ would turn clockwise to $-\pi/2$. 
The phase-amplitude dispersion of Section II then requires that 
$M_{1/2}(D\rightarrow K\pi)$ is enhanced over 
$M_{3/2}(D\rightarrow K\pi)$ by factor $(m^*/m_D)^2$, where $m*$ is 
the highest energy up to which one is willing to make the identification 
of the decay phase with the scattering phase. This relative enhancement 
factor is far too large unless $m*$ is chosen to be comparable to $m_D$.
 
To summarize, the D-meson mass is not low enough to allow identification 
of the final-interaction phase with the elastic scattering phase. 
Yet it is not high enough for our random S-matrix approximation to 
work with a good accuracy.  The large phase difference obtained
by experimentalists can be accommodated but not predicted. 

\subsection{The B decay}

The meson-meson scattering at energy $\sqrt{s} = m_B$ is most likely 
asymptotic for $\pi\pi$, $K\pi$, $\overline{D}\pi$ or even for
$\overline{D}D_s$.  The resonances exist only at much lower energies, 
which means that the contribution of the nondiffractive scattering, 
{\it i.e.,} the non-Pomeron exchanges, to the asymptotic scattering 
amplitude is negligible. For instance, if we make an estimate for
the elastic $K\pi$ scattering at energy $m_B$ using the Regge 
parametrization of Ref\cite{Weyers}, the diffractive contribution gives
\begin{equation}
     |1+2ia_I| = 0.64\;\;(I=1/2,3/2).
\end{equation} 
The nondiffractive contributions modify it only slightly into 
$|1+2ia_{1/2}|=0.63$ and $|1+2ia_{3/2}|=0.65$.
The two-body scattering is clearly in the asymptotic energy region. 
The $s$-wave inelasticity for $K\pi$ is
\begin{equation}
       \sigma_{inel}/\sigma_{tot} = 0.83
\end{equation}
from our estimate in Eq.(\ref{estimate}). There is no chance for the elastic 
approximation to work at this energy. The fact that the largest branching 
fractions observed in the $B$ decay are at the level of 1\% is a clear 
evidence for the presence of very many open channels. The $B$ decay is the 
process where our results of Section IV should apply best. We shall restate 
our predictions for the $B$ decay below.

First of all, the two-body $B$ decay amplitudes should be dominantly 
real up to CP violation phases. According to our very crude estimate made 
in Eq.(\ref{deltaestimate}), the decay phases should be $7^{\circ}$ or 
smaller. The relative phases should be even smaller. As has already been 
pointed out, this is in a sharp contrast to the conclusion of the elastic 
approximation which predicts that the phase differences are small but 
the phases themselves are close to $90^{\circ}$. Though some of the 
existing literature reach the same or similar conclusions, our $s$-channel 
picture for the origin of the almost real decay amplitudes is quite different 
from and orthogonal to them. Our phase-amplitude dispersion relation 
corroborates the smallness of the decay phases. Up to the short-distance
QCD corrections, the magnitude of the final-state interaction phase is 
independent of isospins in the first-order approximation. This small 
and common final-state interaction phase is the reason why the color 
suppression appears to hold well in the $B$-decay.

As for a phase difference between a pair of decay amplitudes, 
the portion of the relative phase that arises from the first-order 
correction to the random phase limit is not calculable, but small 
when the channel number $n$ is large. The typical branching fraction
for the two-body modes $B\rightarrow \overline{D}M, \overline{D}^*M\cdots$, 
where $M$ is $\pi, \rho,\cdots$, is somewhere between 0.1\% and 1\% while the 
inclusive nonleptonic branching due to $b\rightarrow c\overline{u}d$ should 
be close to 70\%.  Therefore the effective number of open channels $n$ with 
this set of quantum numbers is a hundred or more. For the decay modes such as 
$B\rightarrow D^-\rho^+$ that have relatively large branching 
fractions, the randomness approximation gives reliable predictions
while the phases of the suppressed decay modes such as
$B^0\rightarrow\overline{D}^0\rho^0$ cannot be reliably predicted in 
this approximation. In other words, the phase of the $B^0\rightarrow
\overline{D}^0\rho^0$ amplitude can be large. The reason is that the 
two isospin amplitudes $(\overline{D}\rho)_{1/2,3/2}$ of $B^0\rightarrow
\overline{D}^0\rho^0$ nearly cancel each other in the $B^0\rightarrow 
\overline{D}^0\rho^0$ combination, thus enhancing the phase fluctuation 
contributions.

The relative phase which arises from the interference with the subdominant 
nondiffractive eigenchannels is negligibly small since the contribution of
the nondiffractive channels is suppressed by $f_B/m_B$. According to our 
estimate in Eq.(\ref{phasedifference}), the phase difference of this origin 
is expected to be
\begin{equation}
     \Delta\delta \sim 1/300.
\end{equation}
The phase difference due to the fluctuations is more important in 
comparison. We are unable to estimate the relative importance between 
the fluctuations and the short-distance effects to the phase differences. 
Notable exceptions are the decay modes for which the annihilation 
or exchange process dominates. For example, the decay 
$B^0\rightarrow K^+K^-$ for which no spectator diagram can be drawn. 
In this case the decay proceeds only through $B\rightarrow R
\rightarrow K^+K^-$ in the picture of the dual resonance model.  
There are no diffractive eigenchannels in the $s$-channel for this mode. 
Therefore the $B^0\rightarrow K^+K^-$ amplitude can have a large phase 
unlike the amplitude for the spectator-dominated mode 
$B^0\rightarrow K^0\overline{K}^0$. The rescattering from 
$K^0\overline{K}^0$ to the $K^+K^-$ final state can occur through
a $q\overline{q}$ pair exchange, namely, the non-Pomeron exchange.
By stretching the corresponding quark diagram out, however, we find 
that the this rescattering decay process is actually due to one of the 
annihilation diagrams.  Unfortunately the branching fraction 
is too small for such interesting decay modes.

\section{Concluding remarks}
   A reliable computation of the long-distance final-state interaction 
phases is nearly an impossible task. The elastic rescattering 
approximation is probably not viable even for the D decay. It is out of 
question for the B decay. Nonetheless we are able to extract a few 
relevant pieces of information with the phase-magnitude dispersion 
relation and with the eigenphase analysis. The conclusion from the 
dispersion relation is rigorous while the argument based on the 
eigenphase shifts resorts to the random channel-mixing postulate 
and the dual resonance model. The complexity of long-distance strong 
interactions appears formidable if we try to go any step further 
along this line. It may be possible to reach essentially the same 
qualitative conclusions with more an intuitive qualitative 
reasoning\cite{Bjorken}. We have tried in a way different from anybody 
else, maybe in a hard way. We hope that our approach sheds a light 
on some aspects of the problem which have so far not been appreciated.  

\acknowledgements

This work was supported in part by the Director, Office of Energy Research, 
Office of High Energy and Nuclear Physics, Division of High Energy Physics 
of the U.S. Department of Energy under Contract DE-AC03-76SF00098 and in
part by the National Science Foundation under Grant PHY-95-14797.

\noindent
\begin{figure}
\epsfig{file=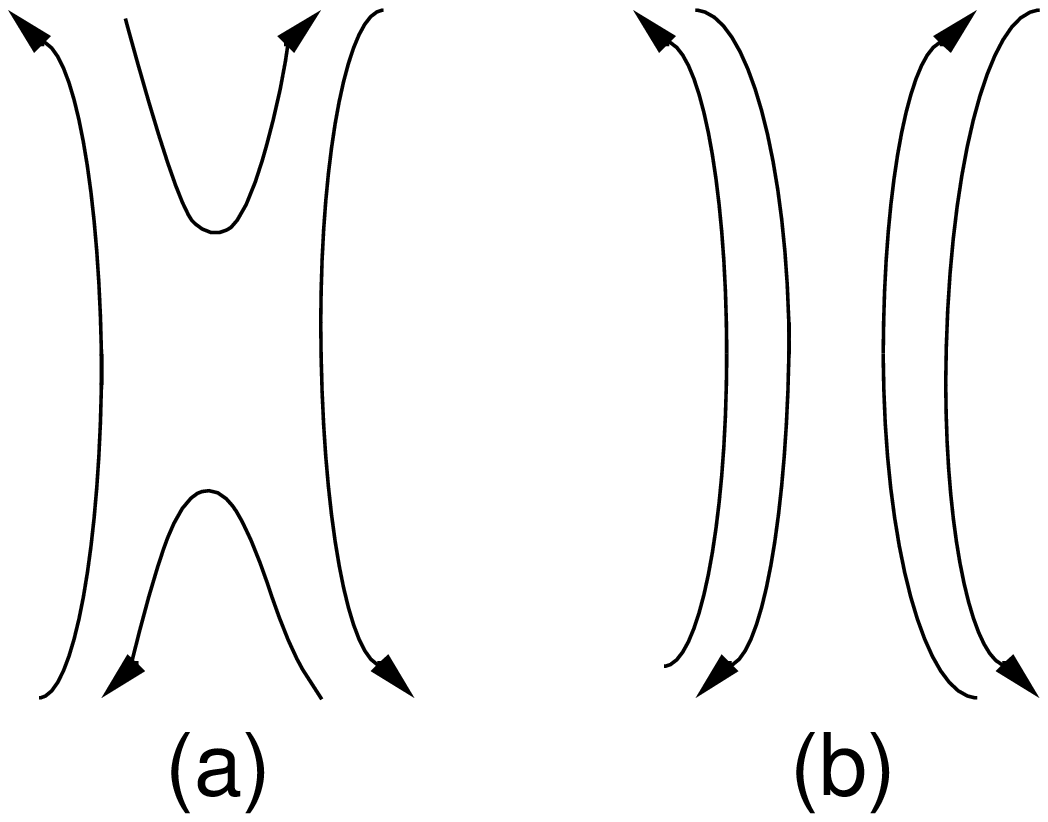,width=7cm,height=3cm}
\vskip 0.5cm
\caption{Duality in the quark diagrams for the elastic boson-boson scattering.
(a)  The resonant channels dual to the non-Pomeron Regge trajectories.
Being made of a $q\overline{q}$ pair, the intermediate resonant states
cannot have the exotic (non-$q\overline{q}$) quantum numbers.
(b) The nonresonant channels dual to the Pomeron consist of
$q\overline{q}q\overline{q}$. In the Pomeron exchange process, one boson 
($q\overline{q}$) exchanges gluons with the other boson ($q\overline{q}$). 
Since gluons carry no flavors, the Pomeron is necessarily a flavor singlet.
The $s$-channel intermediate states consist of a pair of hadrons or 
hadron resonances which does not resonante.}
\label{fig:1}
\end{figure}

\end{document}